\documentclass[%
 reprint,
 amsmath,amssymb,
 aps,
]{revtex4-2}

\usepackage{graphicx}
\usepackage{xcolor}
\usepackage{dcolumn}
\usepackage{bm}

\newcommand{\vte}{v_{te}}
\newcommand{\tpm}{t_{\rm pm}}

\begin{document}

\preprint{APS/123-QED}

\title{Collisionless dynamo seeds from phase mixing-induced electron slippage}

\author{Istv\'{a}n Pusztai}

\email{pusztai@chalmers.se}
\author{Lise Hanebring}%
\affiliation{Department of Physics and Astronomy, Chalmers University of Technology, Gothenburg 41296, Sweden}
\author{James Juno}
\affiliation{
West Virginia University, Morgantown, 26505, WV, United States \\
Lanyon AI Inc., Princeton, 08540, NJ, United States
}%

\date{\today}

\begin{abstract}
Magnetic fields permeating the Universe on the largest astrophysical scales are thought to result from dynamo amplification in weakly collisional turbulence, but the origin of the seed fields remains an open problem of cosmic magnetogenesis. We identify a kinetic mechanism for magnetic-field generation in initially unmagnetized, collisionless plasmas, arising from phase-mixing-induced braking of spatially varying electron flows. Using analytical theory, fully kinetic Vlasov simulations, and turbulent scaling arguments, we show that this process generates coherent magnetic seed fields on scales far larger than the characteristic kinetic scales of the plasma, with strengths comparable to or exceeding classical Biermann battery estimates. The mechanism requires neither a finite initial magnetic field nor misaligned thermodynamic gradients and occurs naturally in electron--ion plasmas.

\end{abstract}

\maketitle



Magnetic fields permeate the Universe on all scales up to the largest gravitationally bound structures \cite{Beck_1996,Carilli2002,Subramanian_2016,Govoni_2004,Kulsrud_2008,Brandenburg_2023}, regulating momentum and heat transport \cite{Kunz_2014,Zhuravleva_2019,Komarov_2016,RobergClark_2016,RobergClark_2018} and cosmic-ray propagation, and playing an important role in particle acceleration \cite{Matthews_2020,Amano_2022,Guo_2024,Sironi_2025} as well as the formation of galaxies and stars \cite{KrumholzFederrath2019,Hopkins2020}. Understanding the origin of cosmic magnetization therefore remains a fundamental problem.

Observations indicate magnetic field energy densities in galaxies and clusters in near equipartition with the kinetic energy in their turbulent flows. Thus these fields are believed to result from turbulent dynamo amplification operating in weakly collisional astrophysical plasmas \cite{Bonafede_2010,Feretti_2012,Crutcher_2012}. Dynamo action, however, requires a pre-existing seed field  \cite{Kulsrud_2008,Brandenburg_2005,Rincon_2019}, the origin of which remains debated. Proposed mechanisms include primordial magnetogenesis in the early Universe, magnetic fields generated during cosmological phase transitions, the Biermann battery \cite{Biermann_1950} operating in ionization fronts. However, the resulting field amplitudes are typically weak \cite{Durrive_2015,Subramanian_2016,Langer_2018}, motivating the search for additional sources of coherent seed fields.

The Weibel instability \cite{Weibel_1959} provides an alternative route to magnetic-field generation \cite{Zhou2022,Zhou_2024,Pucci_2021,Sironi_2023}, efficiently converting free energy associated with pressure anisotropy into magnetic fields and rapidly magnetizing initially unmagnetized plasmas \cite{Medvedev_1999}. However, the resulting fields are generated on kinetic scales, and their relevance for large-scale magnetogenesis depends on the efficiency of subsequent transfer of magnetic energy to larger scales \cite{Liu_2025}. While recent kinetic studies have demonstrated substantial scale growth of Weibel-generated fields, the extent to which such fields can provide large-scale dynamo seeds remains an active area of investigation.

At the same time, numerical studies of collisionless astrophysical dynamos have undergone rapid development \cite{Rincon_2016,StOnge_2018,Zhou_2024,pusztaiPRL,Hanebring_2026}, revealing rich multiscale processes absent from conventional magnetohydrodynamic modeling \cite{Schubert_2011,Tobias_2021,Kapyla_2023,Schekochihin_2004,Porter_2015,Vazza_2017,KorpiLagg_2024} and potentially contributing to magnetic-field generation. Most such studies have employed hybrid-kinetic models or electron--positron plasmas, while fully kinetic simulations of electron--ion plasmas are only now becoming computationally accessible \cite{Bacchini_2026}.

In this Letter we identify a kinetic mechanism for magnetic-field generation in initially unmagnetized electron--ion plasmas. The mechanism relies on phase mixing of spatially varying electron flows. Although electrons and ions are initially current-neutral, phase mixing reduces the electron flow while the ion flow remains nearly unchanged, producing a current that self-consistently generates magnetic fields. Using analytical theory, fully kinetic Vlasov simulations, and turbulent scaling arguments, we show that the resulting fields constitute a competitive source of large-scale magnetic seeds and may play a role both in cosmic magnetogenesis and in the dynamics of collisionless electron--ion plasmas.

We consider magnetic-field generation in an initially unmagnetized, collisionless electron--proton plasma, with both species sharing the initial bulk flow $V_y(x,0)=V_0\exp(ikx)$. The plasma is initially quasineutral, and the identical species flows produce no net current; we therefore take the electromagnetic fields to vanish at $t=0$. For notational convenience, the derivation is carried out using the standard complex-exponential representation, with physical quantities obtained by taking the real part. The characteristic timescale of the process is the electron phase mixing time $t_{\rm pm}= (k\vte)^{-1}$ -- with $\vte=\sqrt{2T_e/m_e}$, $T_e$ and $m_e$ denoting the thermal speed, the temperature and the mass of electrons --  which allows us to disregard the much slower ion dynamics. Furthermore, we expand the near-Maxwellian electron distribution $f_e(\mathbf{x},\mathbf{v},t)$ in $V_0/\vte \ll 1$ for the assumed subsonic flow velocities as $f_e\approx f_0+f_1$, where $f_0=n/(\sqrt{\pi}\vte)^3 \exp[-(v/\vte)^2]$, $n$ is the electron density and $f_1(t=0)=(2V_0v_y/v_{te}^2) \exp(ikx) f_0$. 

To obtain the early-time evolution of the electron current, we consider the limit in which the self-generated magnetic field is sufficiently weak that its effect on particle trajectories can be neglected. With the notation $X=\hat{X}\exp(ikx)$, and $e$ denoting the elementary charge, the electron Vlasov equation linearized about $f_0$ reads  
\begin{equation}
    \partial_t \hat{f}_{1}+ikv_x \hat{f}_{1}=\frac{e}{m_e}\hat{E}_y(t)\frac{\partial f_{0}}{\partial v_y}.
\end{equation} 
Then, similar to a Landau damping calculation we solve the initial value problem using the one-sided Laplace transform $\tilde{X}(\omega)=\int_0^\infty dt \exp(i\omega t) \hat{X}(t)$. This allows us to obtain an algebraic equation, from which $\tilde{f}_1$ can be expressed as   
\begin{equation}
    \tilde{f}_1=i\frac{2 V_0 v_y f_0/v_{te}^2}{\omega-kv_x}+i\frac{e}{m_e}\frac{\partial_{v_y}f_0}{\omega-kv_x}\tilde{E}_y(\omega).
\end{equation}
The current arises from phase-mixing-induced slippage between the electron and ion flows, so that $\hat j_y(t)=enV_0-e\int v_y\hat f_1\,d^3v$. The Laplace transform and the velocity integrals can be carried out in terms of the plasma dispersion function $Z(\zeta)=\pi^{-1/2}\int_{-\infty}^\infty dt e^{-t^2}/(t-\zeta)$, to find
\begin{equation}
    \tilde{j}_y(\omega)=i\frac{e n V_0}{\omega}[1+\zeta Z (\zeta)]-\frac{i\epsilon_0\omega_{pe}^2}{kv_{te}}Z(\zeta)\tilde{E}_y(\omega),
    \label{jytilde}
\end{equation}
where $\zeta=\omega/(k \vte)$, and $\omega_{pe}=\sqrt{ne^2/(\epsilon_0 m_e)}$ is the electron plasma frequency with $\epsilon_0$ the vacuum permittivity. 

For the nonrelativistic astrophysical plasmas of interest, the characteristic electron phase-mixing frequencies satisfy \(\omega\sim kv_{te}\ll kc\), justifying the neglect of the displacement current to relative order $(v_{te}/c)^2$. Combining Ampère's and Faraday's laws yields $\tilde E_y=(i\omega\mu_0/k^2)\tilde j_y$, which provides the electromagnetic closure of the problem by relating the induced electric field to the current density. Although the effect of the self-generated magnetic field on particle trajectories has been neglected, its role in determining the electromagnetic response is retained through this relation. Finally, introducing the electron inertial length $\delta_e=c/\omega_{pe}$, we invert the transform to obtain the time-dependent current amplitude 
\begin{equation}
    \hat{j}_y(t)=\frac{enV_0}{2\pi}\int_{-\infty}^\infty d\omega e^{-i\omega t} \frac{i}{\omega}\frac{1+\zeta Z(\zeta)}{1-\frac{1}{k^2\delta_e^2}\zeta Z (\zeta)}.
    \label{fullcurrent}
\end{equation}
Although Eq.~(\ref{fullcurrent}) can be evaluated numerically, a useful approximation follows from its low-frequency limit $|\omega|\rightarrow 0$, for which $Z(\zeta)\approx i\sqrt{\pi}-2\zeta+\dots$. In the astrophysically relevant regime $k^2\delta_e^2\ll 1$, this yields
\begin{equation}
    \hat{j}_y(t)\approx enV_0[1-e^{-\gamma t}].
    \label{late}
\end{equation}
Equation~(\ref{late}) describes the formal long-time asymptotic behavior of the reduced analytical model, Eq.~(\ref{fullcurrent}), rather than that of the full Vlasov--Maxwell system.

The rate of approach in Eq.~(\ref{late}) is determined by the zero of the denominator of the integrand. In the same limit this gives
$1-i\sqrt{\pi}(k^2\delta_e^2)^{-1}\zeta=0$,
yielding a pole at $\omega=-i\gamma$, with
$\gamma=k^3\delta_e^2v_{te}/\sqrt{\pi}$.
This coincides with the known damping rate \cite{Mikhailovskii_1980,pusztaiPRL}
of magnetic perturbations in non-magnetized collisionless plasmas whose current is carried by electron flows. Here, however, the current is generated by the phase-mixing-induced slippage of electrons relative to the ion flow.

If the system were to continue evolving according to Eq.~(\ref{late}), all electron flow would eventually be phase mixed away, leaving a current carried entirely by the ions with amplitude $enV_0$ and corresponding magnetic field $B_{\rm asy}\sim\mu_0enV_0/k$. However, the approximation underlying Eq.~(\ref{fullcurrent}) ceases to be valid well before this state is reached, because the self-generated magnetic field can no longer be neglected in the particle dynamics. Although the subsequent nonlinear evolution is beyond the scope of the analytical treatment, the breakdown of the approximation provides a natural basis for estimating physically motivated bounds on the eventual saturation amplitude.

 An upper bound for the magnetic field saturation level is given by requiring the field to marginally magnetize thermal electrons at the scale $k^{-1}$, yielding $B_{k\rho_e\sim 1} \sim k v_{te}   m_e/e$.
A lower estimate follows by assuming that saturation
occurs no earlier than one electron phase-mixing time
$t_{\rm pm}$. For $k\delta_e\ll 1$,
Eq.~(\ref{fullcurrent}) gives
$\hat{j}_y(t\sim t_{\rm pm})\sim enV_0 k^2\delta_e^2$,
so Amp\`ere's law yields
\begin{equation}
    B_{t\sim t_{\rm pm}}\sim
    \frac{kV_0m_e}{e}
    =B_{\rm asy}k^2\delta_e^2.
\end{equation}
 We note that, since $B_{k\rho_e\sim1}/B_{t\sim\tpm}\sim\vte/V_0$, on the timescale that would be required for reaching $B_{k\rho_e\sim 1}$ ion phase mixing also becomes non-negligible,  for comparable species temperatures and nearly sonic flow speeds. The analytical theory provides the early-time evolution of the slippage field together with physically motivated bounds on its saturation amplitude. Establishing the nonlinear evolution between these bounds, and determining the resulting saturation level, however, requires first-principles kinetic simulations.

In the following, we employ the Vlasov--Maxwell solver \cite{Juno_2018} of the continuum plasma simulation framework \textsc{Gkeyll} \cite{gkeyllweb} to demonstrate that saturation occurs between the bounds $B_{t\sim t_{pm}}$ and $B_{k\rho_e\sim 1}$, and to verify the expected linear scaling with $k$ in the regime $k\delta_e \ll 1$.  We perform Vlasov simulations of an electron--proton plasma with the physical mass ratio $m_i/m_e \approx 1836$, using one spatial and two velocity dimensions. Both species are initialized with spatially homogeneous Maxwellian distributions and share an initial sinusoidal flow $V_y=V_0\sin(kx)$. The simulation domain is periodic, with size $L$ chosen to accommodate a single wavelength of the perturbation, $k_0 = 2\pi/L$.  Apart from using a non-relativistic formulation of the Vlasov equation for these simulations, the solver does not rely on further approximations.   

In our baseline case, $T_i^{(t=0)}=T_e^{(t=0)}$, $v_{te}/c=0.0626$, and the flow is mildly subsonic, $V_0=0.35\,c_s$, where $c_s=\sqrt{T_e/m_i}$.  The baseline simulation domain is $L = 277.7\,\delta_e$, corresponding to $t_{\rm pm}\,\omega_{pe} = 706.4$. We employ a spatial resolution of $\Delta x = 0.27\,\delta_e$, discretize the $\pm 3$ species thermal speed wide velocity domain using $32$ uniformly spaced cells, and represent the distribution functions within each phase space cell using second-order polynomial basis functions.

\begin{figure}[h]
\includegraphics[width=1.0\columnwidth]{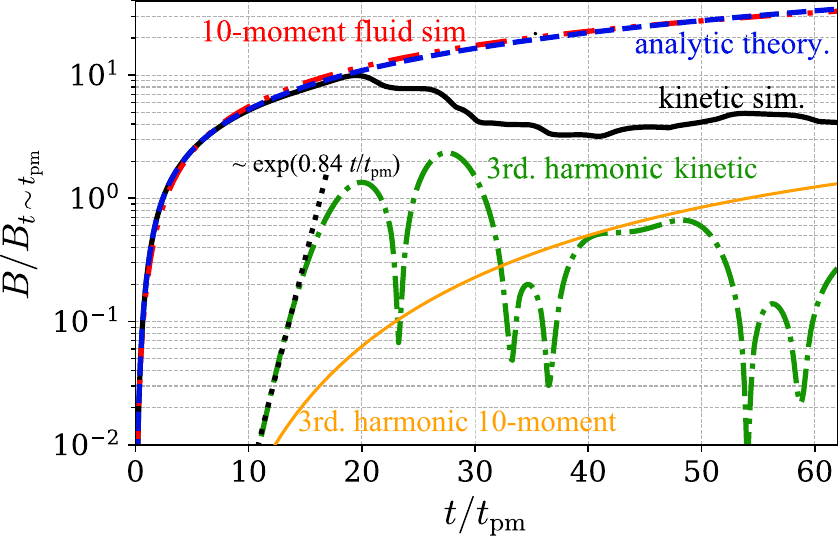}
\caption{\label{fig:timeev} Time evolution of magnetic field amplitude, normalized to $B_{t\sim t_{\rm pm}}$. Solid black: kinetic simulation; blue dashed: linear theory, corresponding to Eq.~(\ref{fullcurrent}); red dash-dot-dot: 10-moment fluid simulation; green dash-dotted: 3$^{\rm rd}$ harmonic in kinetic simulation; thin solid orange: 3$^{\rm rd}$ harmonic in 10-moment simulation; black dotted line shows $\propto \exp(0.84 \, t/t_{\rm pm})$ for reference. The upper-bound estimate $B_{k\rho_e\sim1}\approx 173 B_{t\sim t_{\rm pm}}$ lies above the plotted range.}
\end{figure}

As seen in Fig.~\ref{fig:timeev}, the kinetic simulation result (black solid) closely follows the analytical result (blue dashed) -- given by Eq.~(\ref{fullcurrent}) -- until around $t=19\tpm$, when the kinetic simulation
reaches its maximum amplitude and departs from the analytical prediction.

Preceding the saturation in the kinetic simulation, the odd harmonics of $k_0$ grow exponentially with growth rates $\sim 1/\tpm$. This is illustrated by showing the 3rd harmonic (green dash-dotted) and its fitted exponential growth (black dotted) in Fig.~\ref{fig:timeev}. The harmonics stop growing at the same time the saturation occurs and their total root-mean-square amplitude remains almost an order of magnitude lower than that of the main mode. Although phase mixing generates the pressure anisotropy required to drive the Weibel instability, the electron-Weibel growth rate at the long wavelengths considered here (e.g., the third harmonic) is orders of magnitude too small to explain the observed exponential growth. The selective growth of odd harmonics of the imposed flow wavenumber $k_0$, while even harmonics remain negligible, further distinguishes this process from Weibel instability, which generally amplifies a band of unstable wavenumbers without preferentially selecting odd multiples of $k_0$.

\begin{figure}[h]
\includegraphics[width=1.0\columnwidth]{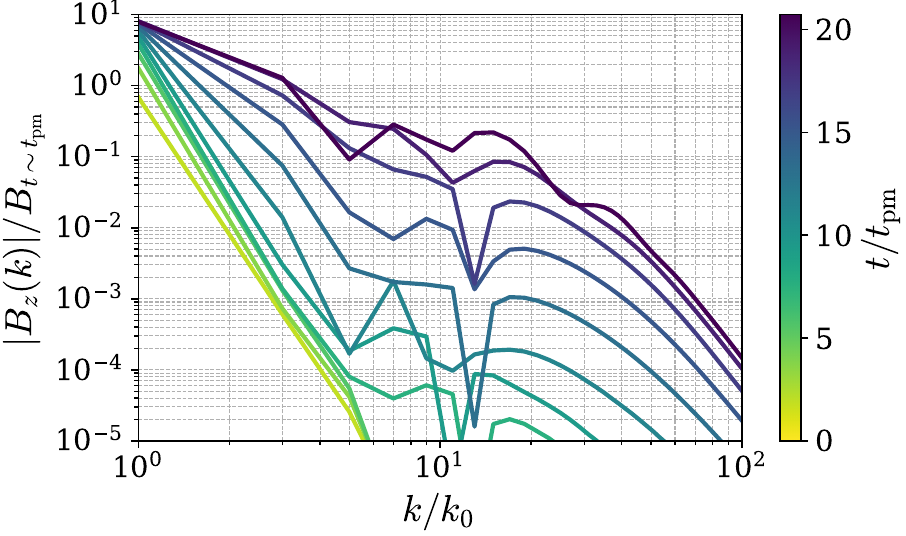}
\caption{\label{fig:spectra} Time evolution of the magnetic field spectrum. Only those wavenumbers that are odd multiples of $k_0=2\pi/L$ are plotted, as even harmonics remain negligible.}
\end{figure}

That the growth of harmonics occurs at a wide range of wavenumbers is illustrated in Fig.~\ref{fig:spectra}, where the time evolution of only the odd harmonics of $k_0$ are shown until the time of saturation. All plotted wavenumbers, besides $k_0$, grow exponentially on $\tpm$ timescales with comparable growth rates. In additional simulations (not shown) with appreciable Weibel growth, the magnetic spectrum instead exhibits a peak near the most unstable Weibel wavenumber. Although nonlinear Weibel evolution can also generate harmonics, the resulting spectral structure retains the breadth of the initial unstable band.     

It is illuminating to model the same scenario using the 10-moment fluid solver \cite{Wang_2015,Wang_2020} of \textsc{Gkeyll}, which evolves the collisionless hierarchy of fluid equations up to the full pressure tensor for each species $\alpha$. Here we use a pressure isotropization heat flux closure \cite{Wang_2015} $ \partial_k q_{ijk,\alpha} = k_{0\alpha} v_{\text{th},\alpha}(p_{ij,\alpha} - p_\alpha \delta_{ij})$ where $q_{ijk,\alpha}$ and $p_{ij,\alpha}$ are elements of the heat-flux and pressure tensors, $p_\alpha$ is the scalar pressure, and $k_{0\alpha}$ is a parameter that sets the isotropization strength. At sufficiently high mass ratio $k_{0i}$ sets the damping rate of the mass flow, while $k_{0e}$ affects the damping rate of the magnetic field when the current is carried by electrons, as described in Ref.~\cite{Hanebring_2026}. The ten-moment model captures sufficient information about the phase-space structure that, similarly to magnetic field damping, it can approximately reproduce the dynamics of the stress developing due to phase-mixing breaking the electron flow, and the rate of the corresponding electron slippage is effectively tuned by $k_{0e}$. When using $k_{0i}/k_{0}=100$ to reduce ion flow damping to negligible values, $k_{0e}/k_{0}=1.1$ yields a close agreement between the magnetic field growth predicted by Eq.~(\ref{fullcurrent}) and the 10-moment simulation; compare the blue dashed and red dash-dot-dot curves in Fig.~\ref{fig:timeev}. Importantly, the fluid result does not saturate during the simulation time and, unlike the kinetic simulation, the growth of odd harmonics is polynomial instead of exponential (see thin orange line in Fig.~\ref{fig:timeev}). This suggests that the saturation observed in the kinetic simulation has an origin involving finer structures in the distribution function. 

\begin{figure}[h]
\includegraphics[width=0.9\columnwidth]{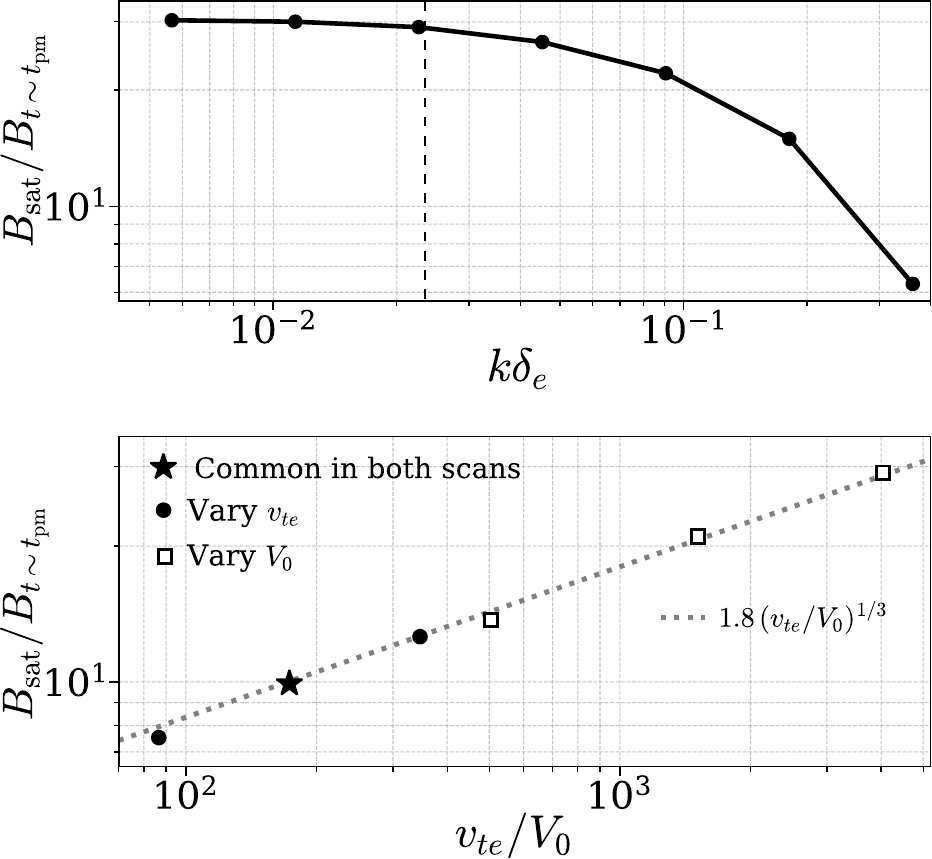}
\caption{\label{fig:kscaling} (a) Normalized simulated saturation levels, $B_{\rm sat}/B_{t\sim t_{\rm pm}}$, showing an approach to a constant value as $k\delta_e$ decreases. The vertical dashed line marks $k\delta_i=1$. (b) Normalized saturation levels from scans varying either $v_{te}$ or $V_0$. The dotted line represents $B_{\rm sat}/B_{t\sim t_{\rm pm}}=1.8(v_{te}/V_0)^{1/3}$, supporting the empirical scaling of Eq.~(\ref{satrule}). The star marks the simulation common to both parameter scans. }
\end{figure}

To determine the wavenumber scaling of the saturation field strength, we perform fully kinetic Vlasov--Maxwell simulations scanning $k\delta_e$, extending down to $k\delta_i\simeq 0.24$, using $V_0=0.015\,c_s$. As shown in Fig.~\ref{fig:kscaling}(a), the normalized saturation level approaches a constant at small $k\delta_e$, consistent with $B_{\rm sat}\propto k$. Additional scans, independently varying $v_{te}$ and $V_0$ at fixed wavenumber, yield the collapse shown in Fig.~\ref{fig:kscaling}(b), supporting the empirical scaling
\begin{equation}
    B_{\rm sat}\sim (v_{te}/V_0)^{1/3} k V_0 m_e/e.
    \label{satrule}
\end{equation}

The saturation observed in the kinetic simulations is closely related to the growth of odd harmonics of the magnetic field, suggesting a nonlinear harmonic-transfer process as a possible saturation mechanism. The Lorentz force associated with the fundamental current and magnetic field contains a component at twice the fundamental wavenumber. This force distorts the current profile and generates a third-harmonic component, leading to a hierarchy of odd magnetic harmonics, while even harmonics remain weak due to the parity of the nonlinear coupling.

A phenomenological estimate of the saturation amplitude can be obtained by assuming an effective nonlinear transfer rate $\gamma_{\rm nl}\sim kV_0(\Omega_e/kV_0)^3$, where $\Omega_e=eB/m_e$ is the electron cyclotron frequency. The cubic dependence is motivated by the multistage coupling within the observed odd-harmonic hierarchy, rather than derived from a controlled perturbative expansion. Assuming that saturation occurs when this transfer rate becomes comparable to the electron phase-mixing rate $kv_{te}$, which also characterizes the observed harmonic growth, recovers the empirical scaling in Eq.~(\ref{satrule}).

Further insight is provided by comparison with the 10-moment model. The fluid simulations reproduce both the linear phase-mixing growth of the fundamental mode and the emergence of odd harmonics. However, the harmonics grow algebraically in time and no saturation is observed at the amplitudes reached in the kinetic simulations. In contrast, the kinetic simulations exhibit an approximately exponential amplification of the odd harmonics on a timescale comparable to the inverse phase-mixing rate. This suggests that the odd-harmonic hierarchy results from nonlinear mode coupling, whereas the rapid transfer through that hierarchy is intrinsically kinetic and likely mediated by the phase-space structure generated by phase mixing.

Although the nonlinear saturation mechanism remains incompletely understood, Eq.~(\ref{satrule}) provides an empirical estimate for the maximum seed field generated by the slippage process.
In a turbulent intracluster medium (ICM), the velocity field spans scales from an outer scale of hundreds of $\rm kpc$ (denoted by the subscript ``out'') down to much smaller scales. To estimate the strength of the slippage seed, we assume a Kolmogorov spectrum,
$
V(k)\sim V_{\rm out}\left(k/k_{\rm out}\right)^{-1/3},
$
together with representative ICM parameters. 
For this velocity spectrum, Eq.~(\ref{satrule}) predicts
$
B\propto kV(k)^{2/3}\propto k^{7/9}.
$
Taking $L_{\rm out}=300\,{\rm kpc}$, $T_e=T_i=10\,{\rm keV}$, and an outer-scale Mach number $M=V_{\rm out}/c_s=0.35$, this scaling yields slippage seed fields of order $10^{-24}\,{\rm T}$ at scales of order $100\,{\rm pc}$, comparable in magnitude to cosmological Biermann-battery estimates \cite{Subramanian_1994,Kulsrud1997,Gnedin2000}, with larger amplitudes at smaller scales. Weibel-generated fields can be much stronger ($\sim10^{-14}\,{\rm T}$ for representative ICM parameters \cite{Hanebring_2026}), but initially emerge on kinetic scales of order $\mu{\rm pc}$ or smaller \cite{Zhou_2024}. In contrast, the slippage mechanism generates weaker seeds directly on flow scales, requiring substantially less inverse magnetic-energy transfer before dynamo amplification.

These estimates should be regarded as illustrative rather than predictive. Both the plasma conditions during the initial magnetization of the ICM and the spectrum of turbulence in initially unmagnetized collisionless plasmas remain uncertain. Nevertheless, the scaling analysis demonstrates that the slippage mechanism can produce magnetic seeds on scales far larger than kinetic plasma instabilities and with amplitudes competitive with other large-scale seed-generation mechanisms.

Pressure-anisotropy-driven microinstabilities may modify this estimate. In magnetized high-$\beta$ plasmas, firehose and mirror fluctuations can scatter or trap particles, interrupt phase mixing, and reduce collisionless damping \cite{Kunz_2014,Kunz2020,Meyrand2019}. Similar effective collisionality in an initially weakly magnetized plasma could reduce the electron slippage that drives the present mechanism. However, these instabilities require pressure anisotropy to develop, which occurs on the same phase-mixing time that generates the lower-bound field $B_{t\sim\tpm}$. Thus, even in the limiting case where microinstabilities rapidly interrupt further phase mixing, fields of order $B_{t\sim\tpm}$ are expected. Moreover, simulations in which Weibel growth is significant still exhibit appreciable slippage-generated fields, and contemporary fully kinetic electron--ion dynamo simulations observe the same effect \cite{Bacchini_2026}.  Extrapolating these results to asymptotically scale-separated astrophysical plasmas remains an open problem. The mechanism does not fundamentally require a large ion-to-electron mass ratio (in fact, it arises already in simulations with $m_i=2m_e$), but rather different phase-mixing rates for the two species. 

\emph{In conclusion}, we have identified a kinetic seed-field generation mechanism that requires only spatially varying flows and arises from the phase-mixing-induced slippage of the electron flow. Unlike the Biermann battery, the mechanism operates even when both the initial magnetic field and misaligned density and temperature gradients ($\nabla T\times\nabla n$) vanish identically. We derived an analytical expression for the early-time evolution of the resulting magnetic field and verified it with fully kinetic Vlasov simulations. These simulations further establish the scaling of the nonlinear saturation amplitude.
While the slippage field is also present in the collisionless 10-moment model, the nonlinear saturation observed in the kinetic simulations is associated with the emergence of odd harmonics after several phase-mixing times and is not reproduced by the fluid model, suggesting that kinetic phase-space structure plays a key role in the saturation process.
In the intracluster medium, the resulting seed fields can become competitive with Biermann-battery estimates while being generated on scales many orders of magnitude larger than those associated with the Weibel instability. Determining the role of this mechanism in astrophysical magnetogenesis will require future kinetic studies of weakly magnetized turbulence with greater scale separation, but the present results establish phase-mixing-induced electron slippage as a viable source of magnetic seed fields in collisionless plasmas.

\emph{Acknowledgments:} The authors are grateful for fruitful discussions with M.~Zhou, F.~Bacchini, P.~Helander, E.~Ingelsten and T.~F\"{u}l\"{o}p. The work was supported by the Swedish Research Council (Dnr.~2021-03943) and the Knut and Alice Wallenberg foundation. The development of \texttt{Gkeyll} were partly funded by the NSF-CSSI program, Award No. 2209471. This work was partially completed while James Juno was at the Princeton Plasma Physics Laboratory (PPPL). While at PPPL his work was supported by the U.S.~Department of Energy under Contract No.~DE-AC02-09CH1146 via LDRD grants. The computations were enabled by resources provided by the National Academic Infrastructure for Supercomputing in Sweden (NAISS), partially funded by the Swedish Research Council through grant agreement No.~2022-06725. The authors also acknowledge the Texas Advanced Computing Center (TACC) at The University of Texas at Austin for providing computational resources that have contributed to the research results reported within this paper.

\bibliography{PhaseMixingSeed}

\end{document}